\documentclass[twocolumn,showpacs,preprintnumbers,amsmath,amssymb,pre]{revtex4}
\usepackage{graphics}
\usepackage{graphicx}
\usepackage{amsmath,amssymb,epsfig,picinpar,graphics,float,amssymb,verbatim}
\newcommand{\eq}[1]{Eq.\ (\ref{#1})}

\newcommand{\rem}[1]{}
\newcommand{\dt}[1]{\dot{#1}}
\newcommand{\nsum}[1]{\langle#1\rangle}

\newcommand{\refeq}[1]{Eq.\ (\ref{#1})}
\newcommand{\reffig}[1]{Fig.\ \ref{#1}}

\begin{document}

\title{Coherent regimes of mutually coupled Chua's circuits}
\author{I. Gomes Da Silva}
\thanks{iacyel@imedea.uib.es}
\affiliation{Instituto Mediterr\'{a}neo de Estudios Avanzados
IMEDEA (CSIC-UIB), Campus Universitat Illes Balears, E-07122 Palma
de Mallorca, Spain}
\author{S. De Monte}
\thanks{demonte@biologie.ens.fr}
\affiliation{Dept. of Biology, UMR 7625, Ecole Normale Sup\'erieure,
  Paris, France}
\author{F. d'Ovidio}
\thanks{Now at LMD, Ecole Normale Sup\'erieure, Paris, France, dovidio@lmd.ens.fr}
\affiliation{Instituto Mediterr\'{a}neo de Estudios Avanzados
IMEDEA (CSIC-UIB), Campus Universitat Illes Balears, E-07122 Palma
de Mallorca, Spain}
\author{R. Toral}
\thanks{raul@imedea.uib.es}
\affiliation{Instituto Mediterr\'{a}neo de Estudios Avanzados
IMEDEA (CSIC-UIB), Campus Universitat Illes Balears, E-07122 Palma
de Mallorca, Spain}
\date{\today}
\author{C. R. Mirasso}
\thanks{claudio@galiota.uib.es}
\affiliation{Departament de F\'{\i}sica, Universitat de les Illes
Baleares, Campus Universitat Illes Balears, E-07122 Palma de
Mallorca, Spain}
\date{\today}

\begin{abstract}
We study the dynamical regimes that emerge from the strongly
coupling between two Chua's circuits with parameters mismatch. For
the region around the perfect synchronous state we show how to
combine parameter diversity and coupling in order to robustly and
precisely target a desired regime. This target process allows us
to obtain regimes that may lie outside parameter ranges accessible
for any isolated circuit. The results are obtained by following a
recently developed theoretical technique, the order parameter
expansion , and are verified both by numerical simulations and on
electronic circuits. The theoretical results indicate that the
same predictable change in the collective dynamics can be obtained
for large populations of strongly coupled circuits with parameter
mismatches.
\end{abstract}

\maketitle

\section{Introduction}


One of the most important features of interacting oscillating
units is the possibility for their motions to entrain each other
and evolve in synchronicity. The phenomenon of synchronization is
a universal feature underlying the collective behavior of
populations of dynamical systems and its importance has been
recognized in many different fields
\cite{pikovsky03,mosekilde02,mikhailov04}. In particular, a great
deal of attention has been devoted to globally-coupled dynamical
systems as a model for a variety of physical \cite{oliva01},
chemical \cite{kuramotobook,kiss02c} and biological systems
\cite{dano99,winfree01}.

Contrary to the assumption of identical units, common in many
theoretical approaches, real systems display an unavoidable
diversity in the oscillator population showing up as, e.g.,
parameter mismatches. The effect of such a diversity might hinder
synchronization and yields an incoherent regime if the coupling is
too weak. The main feature of incoherence is that the motions of
the individual systems are largely independent to each others. The
collective dynamics, then, is either stationary in average (for
infinite number of elements) or displays fluctuations that scale
with the population size. When the coupling strength is increased,
some or all the elements of the population synchronize and, as a
result, oscillations start to be detected at a macroscopic level.
The transition to collective oscillations is typically accompanied
by rather complex regimes, such as clustering \cite{osipov98},
partial synchronization \cite{matthews91}, phase synchronization
\cite{rosenblum96}, etc. For even stronger coupling, complete
synchronous regimes, characterized by a high degree of coherency
within the population, develops. This means that all the
population elements share the same kind of dynamics, apart from
small differences due to the microscopic diversity, and this
dynamics reflects at the macroscopic level.

Despite of the numerous theoretical advances in the analysis of
globally coupled dynamical systems, the experimental verification
has revealed to be extremely difficult, in particular as far as
biological systems are concerned. The Kuramoto transition from
incoherence to the locked state is in this sense a good example
\cite{kuramoto75}. It dates back to 1975 and is one of the best
known and general results in synchronization theory. However, to
our  best knowledge, it has been quantitatively verified once in
experiments and required the implementation of an ad hoc system
\cite{kiss02}. Two main problems are encountered in experiments.
One is the difficulty in controlling or measuring the parameter
variability in the system. The second one is related to the global
coupling which is very rare in nature. These experimental
limitations are especially relevant for regimes close to the
incoherence one where a large population is needed.

In order to avoid these problems associated with incoherence, in
this paper we focus instead on globally coupled systems close to
the \emph{synchronous} state. As it was recently shown, such
\emph{coherent} states do not strongly depend on the population
size \cite{demonte02,demonte03}. This fact is very interesting
from the experimental viewpoint because the main characteristics
can be studied in systems of only two elements and the results
extended to larger populations. Moreover, the coherent dynamics
may be described by a set of low dimensional equations with only
few parameters. Transitions among different regimes correspond to
bifurcations involving a small number of dimensions and
parameters, thus having the advantage of being very robust to
changes in the microscopic features of the population.

As an application of such theoretical results, we study the strong
coupling regimes of two Chua's circuits. Chua's circuits offer an
intermediate step between theory and applications. They can be
modelled with a simple set of non-linear equations, thus allowing
the use of analytical methods and quick numerical simulations. At
the same time they can be easily implemented in hardware. Even if
they constitute a rather controlled experimental system, a
verification of theoretical results faces a test on its robustness
with respect to: (i) noise fluctuations in states and parameter
values, (ii) constrains in the accessible parameter precision and
ranges, and (iii) small functional mismatches between model and
implementation. Finally, Chua's circuits are particularly
interesting because they constitute prototypical devices for
nonlinear system testbeds.

Section\ \ref{sec:chuaope} derives the equations for the
macroscopic dynamics of an interacting population (of any size) in
the coherent regime. This representation allows us to infer on the
dynamical regimes of the experimental system. We then provide two
examples of comparison between theoretical prediction and
experimental system. In Sec.\ \ref{sec:strongcoup} we show how, by
controlling parameter mismatch, one can drive the mean field to
regimes that are different from the uncoupled dynamics and that
might not be accessible by any of the individual systems. In Sec.\
\ref{sec:oscdeath} we consider the case in which the individual
elements differ for the time scale of their motion, and focus in
particular on the regime of oscillaton death, where parameter
diversity induces a suppression of the individual and collective
oscillations. Finally, Sec.\ \ref{sec:disc} is devoted to the
discussion of the results and to their possible applications.

\section{Order parameter expansion of mutually coupled Chua's circuits}
\label{sec:chuaope}

The order parameter expansion \cite{demonte02,demonte03} is a
useful technique for studying regimes that bifurcate from the
perfectly synchronous state in populations of globally and
strongly coupled elements when the parameter mismatch is
increased. Let us here review the main results by considering a
system of the form:
\begin{equation}\label{eq:pxj}
\dt{x_j}=f\left(x_j,p_j\right)+K(X-x_j)
\hspace{10mm}j=1,2,\dots,N.
\end{equation}
where $p_j$ is a parameter containing the diversity of the system
and the dynamics of the each element
$f_j(x_j)=f\left(x_j,p_j\right)$, is defined by the smooth
function $f_j: \mathbb R ^n \rightarrow \mathbb R ^n$. All the
elements are coupled to the the mean field of the population
$X=\nsum{x_j}$ through the coupling function $K(X-x_j)$ that, as
we will see later, for the case of the electronic coupling we use,
can be assumed as a linear diagonal operator (in fact, a more
general system can be considered, see \cite{demonte04a} for
details.) The basic idea of the method is to obtain an effective
equation of motion for the mean field variable $X$, valid when all
the elements evolve in time close to $X$. This is done by
considering regimes close to the perfectly synchronous state,
defined as $x_j=X, \, \forall j$. Under this assumption, equation\
(\ref{eq:pxj}) can be expanded into the individual deviations from
the mean field $\epsilon_j=x_j-X$ and from the mean parameter
$\delta_j=p_j-p_0$, being $p_0=\nsum{p}$. By averaging
\refeq{eq:pxj} over the population, one gets, at first order, the
following approximated equation for the mean field:
\begin{eqnarray}\label{eq:pope} \begin{cases}
\dt{X}=F(X,p_0)+D_{x,p}F(X,p_0)\,W\\
\\ \dt{W}=\sigma^2\,D_{p}\,F(X,p_0)+(D_{x}\,F(X,p_0)-K)\,W,
\end{cases}
\end{eqnarray}
where $W=\nsum{\delta\,\epsilon}$ is a second macroscopic
variable, or shape order parameter; $p_0$ is the average parameter
value and $\sigma^2$ is the standard deviation of the parameter
distribution. Besides being low dimensional, such equation has the
advantage of having a clear physical meaning: the mean field
behaves like the average element, perturbed by a macroscopic
variable that quantifies the mismatch in phase and parameter
space. The macroscopic regimes thus appear as an \emph{unfolding}
of the perfectly locked state $\dt{X}=F(X,p_0)$ and are controlled
by the parameter $K$ and $\sigma^2$. Such equation holds when the
displacements from the mean field $\epsilon_j$ are small, and
thus, operationally, when the coupling is sufficiently strong
compared with the parameter mismatch. Another important
consideration is that \refeq{eq:pope} does not depend on the
population size. Consequently, the results we obtain for two
Chua's circuits can be readily generalized to larger arrays.
\begin{figure}[h] \center
\epsfig{file=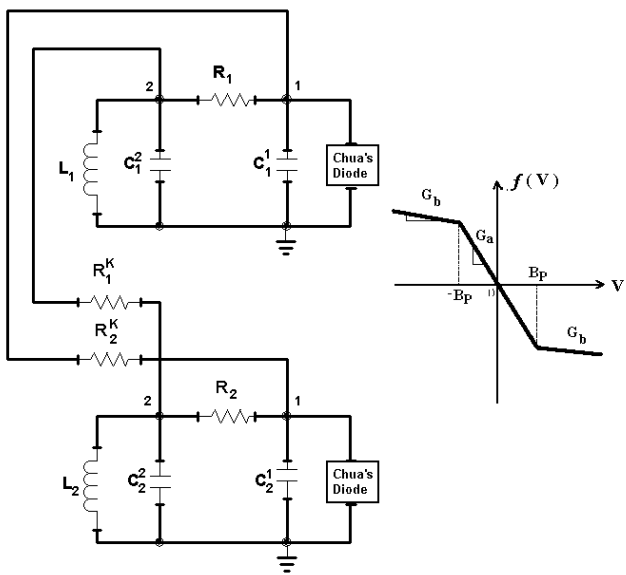,clip=,width=.48\textwidth} \caption{Scheme
of the two mutually coupled Chua's circuits. The parameters are
$C_{1,2}^1=10.2$ nF ,$C_{1,2}^2=101.8$ nF, $L_{1,2}=18.4$ mH,
$R_{1,2}=1562$ Ohm, $G_{1,2}^{a}=-7.625\,10^{-4}$,
$G_{1,2}^{b}=4.12\,10^{-4}$.\label{fig:chua}}
\end{figure}

Let us now apply \refeq{eq:pope} to the case of two mutually
coupled Chua's circuits. The configuration of the circuits that we
study is shown in \reffig{fig:chua}. The equations are easily
written in terms of the voltages $v_1$ and $v_2$ (at point 1 and 2
respectively) and the current $I$ at point 3, the first circuit is
described by:
\begin{eqnarray}\label{eq:chua} \begin{cases}
\dt{v}_1^1=&\frac{1}{C_1^1 R_1}\,v_1^2
-\frac{G_1^a-G_1^b}{2C_1^1}\left(|v_1^{1}+B_1^p|-|v_1^{1}-B_1^p|\right)\\
\rule[-6mm]{0mm}{11mm}
&-\frac{1}{C_1^1}\left(G_1^b+\frac{1}{R_1}\right) v_1^1
+\frac{1}{C_1^1 R^k_1}\, (v_1^1-v_2^1\\ \rule[-6mm]{0mm}{8mm}
\dt{v}_1^2=&\frac{1}{C_1^2}\,I^1+\frac{1}{C_1^2
R_1}\left(v_1^1-v_1^2\right) +\frac{1}{C_1^2 R^k_2} \,(v_1^2-v_2^2)\\
\rule[-3mm]{0mm}{8mm} \dt{I}_1=&-\frac{r_1^b}{L_1}\,
I_1-\frac{1}{L_1}\,v_1^2
 \end{cases}
\end{eqnarray}
where  $R^{1,2}_k$ are the two coupling resistors. These resistors
will be used as control parameters. Other internal parameters,
that are kept constant, are displayed in the caption of Fig.\
\ref{fig:chua}. The second circuit has an equivalent equation,
with lower index 2 instead. Before proceeding with the analysis,
it is worth mentioning that the mutual coupling \refeq{eq:chua}
term can be recast in the form of a global coupling term, so that
the correspondence with \refeq{eq:pope} is readily established.
The dependence on the mean field becomes apparent by noticing
that:
\begin{equation}
\left(v_2^1-v_1^1\right)=\left(v_2^1+v_1^1-2v_1^1\right) =2 \left(
\nsum{v_1}-v_1^1\right).
\end{equation}
The coupling constants are the multiplicative factors $1/(C^1
R^1_k)$ and $1/(C^2 R^2_k)$. At this point, we recover
\refeq{eq:pxj} by identifying $x_j$ with the vector ${<v_j^1, \,
v_j^2,\, I_j>}$ and $f_j(x_j)$ with the equation for an
\emph{uncoupled} Chua circuit. The coupling matrix is:
\begin{eqnarray}
K^1=\left[
\begin{matrix}
\frac{1}{C_1^1 R^1_k} & 0 & 0\\
0 & \frac{1}{C_1^2 R^2_k} & 0\\
0 & 0 & 0
\end{matrix}
\right]
\end{eqnarray}

Since we assume identical capacities for the two circuits, the
coupling is symmetrical. The explicit calculation of the reduced
equations are given in the Appendix. With this formalism, we are
now ready to use \refeq{eq:pope} on some specific condition and to
study the collective regimes that emerge due to the interplay of
parameter mismatch and coupling.

\section{Strong mutual coupling}
\label{sec:strongcoup}

Let us first address the case in which the mismatch is in the
internal resistors $R_{1,2}$ of the two circuits, while the other
internal parameters are kept identical. The coupling resistors
$R_k^{1,2}$ are chosen weak enough to ensure that the coupling is
strong and the regime coherent. Under such strong coupling
conditions, the collective dynamics will be qualitatively equal to
the dynamics of an ``average oscillator''. This corresponds to a
system of equations identical to those of each single element,
except for the parameter on which diversity is imposed, that is
instead substituted by its average value. Intuitively, such
property of the macroscopic dynamics is due to the fact, that when
the coupling matrix is predominantly diagonal, the shape parameter
$W$ is kept small. This corresponds to the situation in which the
population is tightly packed around the mean field. In the limit
in which the coupling is infinitely strong, $W$ vanishes and in
\refeq{eq:pope} we end up with the equation for an average
oscillator.

Such a consideration has interesting implications in the control
of the collective behavior of populations of dynamical systems, as
well as in the detection of the microscopic features in terms of
purely macroscopic observations. Indeed, the dynamics that is
stabilized by coupling non-identical circuits may not be
attainable by any of them if uncoupled. The collective regime is
thus ``coded'' in the diversity and coupling rather than in the
internal features of the individual circuits. Moreover,
\refeq{eq:pope} allows a straightforward identification of such a
hidden collective behavior.

\begin{figure}[h] \center
\epsfig{file=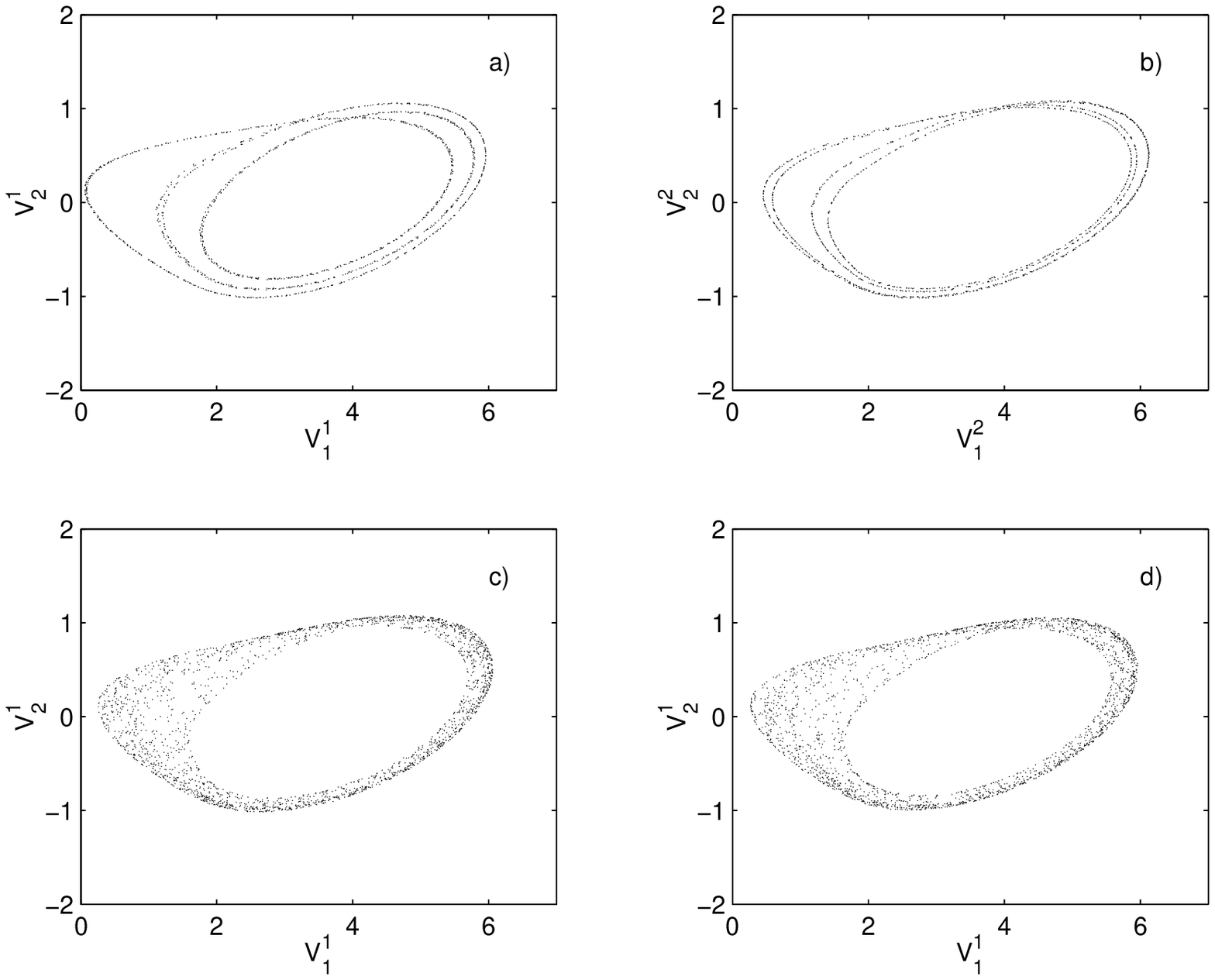,clip=,width=.48\textwidth}
\caption{Collective chaotic behavior as generated by coupling
  two nonidentical circuits, each of which is individually
  periodic. a) and b) Uncoupled chaotic dynamics
  of the two circuits ($R_1=1740$ Ohm, $R_2=1644$ Ohm). c) Attractor of
  the first circuit after the coupling is established.
  Being the coupling strong ($R_k^{1,2}=200$ Ohm), the attractors
  of the second circuit and of the mean field are nearly identical to
  this one and thus are not shown. d) The same behavior of the strongly coupled
  circuits is predicted by a single circuit with the
  internal resistance set at $(R_1+R_2)/2=1692$ Ohm.
  \label{fig:cyc_cyc}}
\end{figure}

As a first application of the aforementioned theoretical results,
let us examine the case when both circuits operate in a periodic
regime. Figure\ \ref{fig:cyc_cyc} (a) and (b) displays the
experimentally measured attractors of the two uncoupled circuits,
of period 3 and 4 respectively. Figure\ \ref{fig:cyc_cyc} (c) and
(d) show that, as a result of the coupling, both oscillators
behave chaotically, although the mismatch in the parameters has
only minor effects on their individual dynamics.

\begin{figure}[h] \center
\epsfig{file=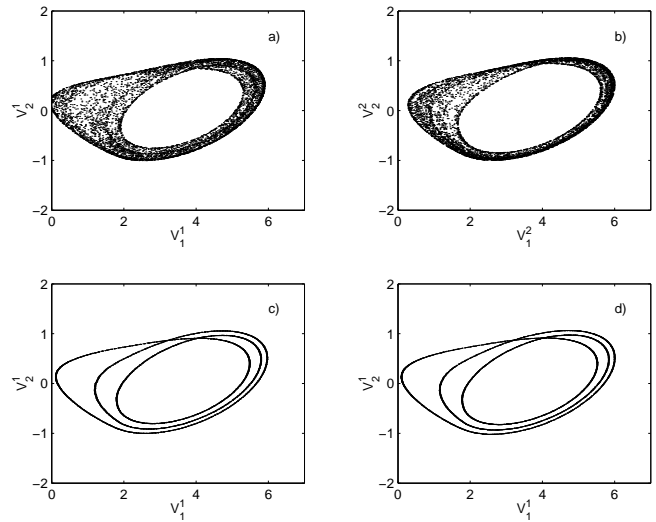,clip=,width=.48\textwidth}
\caption{Stabilization of an unstable period-three orbit by coupling
  of two nonidentical chaotic circuits. a) and b) Uncoupled chaotic dynamics
  of the two circuits ($R_1=1621,\, R_2=1745$ Ohm). c) Attractor of
  the first circuit after the coupling is established. Being the coupling strong, the attractors
  of the second circuit and of the mean field are nearly identical to
  this one and thus are not shown. d) As in the Fig.\ref{fig:cyc_cyc} the
  dynamics of the coupled circuits can be predicted by a single circuit with the
  internal resistance at the average value $(R_1+R_2)/2=1683$ Ohm.
  \label{fig:chaos_chaos}}
\end{figure}

The same idea can be used for stabilizing unstable periodic
orbits. Let us, for instance, consider the case in which both
circuits are chaotic, but their parameter values are located, when
the circuits are uncoupled, on the opposite side of a periodic
window. Figure\ \ref{fig:chaos_chaos} shows that once they are
coupled, both circuits get synchronized onto a period-three orbit.
The coupling thus results into the stabilization of an unstable
orbit embedded into the chaotic attractor.

The possibility of obtaining a qualitatively novel dynamics by
coupling two circuits with different internal parameters is robust
with respect to changes in the coupling intensity (also under
differential modifications of the resistors) as long as both
coupling resistors are sufficiently weak, that is coupling is
sufficiently strong. This robustness was confirmed by repeating
the two experiments described above for several values of the
coupling resistors.

Besides providing a way for targeting specific regimes, the
experiments also show that the mathematical limit of infinitely
strong coupling is actually attainable, to a good approximation,
even for experimental feasible resistance values.
It is also important to remark that in order to target a regime one only needs
to know the bifurcation diagram: this information can be obtained experimentally
and in principle does not require the knowledge of the functional form of $f$.

\section{Oscillation death in circuits with time scale mismatch}
\label{sec:oscdeath}

In this section we address oscillaton death, a regime where the
dynamics of each population element is suppressed as a collective
effect, i.e., resulting from the interplay of coupling and
diversity. Oscillator death has been firstly described in
populations of limit cycle oscillators with direct
\cite{ermentrout89,ermentrout90} or time-delayed coupling
\cite{atay03b}. Recently, an order parameter expansion has been
used to demonstrate that this regime may arise under generic
conditions in populations of globally coupled element with time
scale mismatch \cite{demonte03}. Such results introduce us to the
next study on coupled circuits. The equations for a population
with time scale mismatch and global, linear coupling can be
written as:
\begin{equation}\label{eq:tauj}
\dt{x_j}=\tau_j f\left(x_j\right)+K(X-x_j),
\hspace{10mm}j=1,2,\dots,N.
\end{equation}
This equation may be seen as a simple way for generalizing the
Kuramoto model, where the distributed parameter is the natural
frequency of the oscillators, to an individual dynamics different
from a limit cycle. The peculiar parameter dependence of
\refeq{eq:tauj} is reflected in a simple form of the reduced
system:
\begin{eqnarray}\label{eq:popetau}
\begin{cases} \dt{X}=\nsum{\tau}f(X)+J\,W\\
\\
\dt{W}=\sigma^2\,f(X)+(\nsum{\tau}J-K)\,W,
 \end{cases} \end{eqnarray}
where $J$ is the Jacobian $D_x f(x)$.

Due to the multiplicative nature of the distributed parameter in
\refeq{eq:tauj}, any fixed point of $f$ is also a fixed point for
the mean field. The stability of such a point, however, can change
due the presence of the coupling and time scale mismatch. With
some algebra one finds that generic conditions exist under which
an unstable focus of $f$ becomes attracting for the population.
Referring to \cite{demonte03} for the details, here we just notice
that, if the time scale mismatch is sufficiently wide, this
equilibrium is stabilized for ``intermediate'' values (i.e., large
enough to avoid incoherence and smaller than those where
nonstationary coherent solutions are present) of the coupling.

For the case of Chua circuits, one can see from \refeq{eq:chua} that the time scale
can be easily changed by tuning the capacitors $C_1$,
$C_2$ and the inductance $L$. We modify the time scale of one of
the two circuits by making use of commercially available
capacitors and inductances, thus obtaining a mismatch of at most
14\%. Both circuits, when uncoupled, have similar chaotic dynamics
(double scroll attractor) although their time evolution takes
place with different speed.

\begin{figure} \center
  \epsfig{file=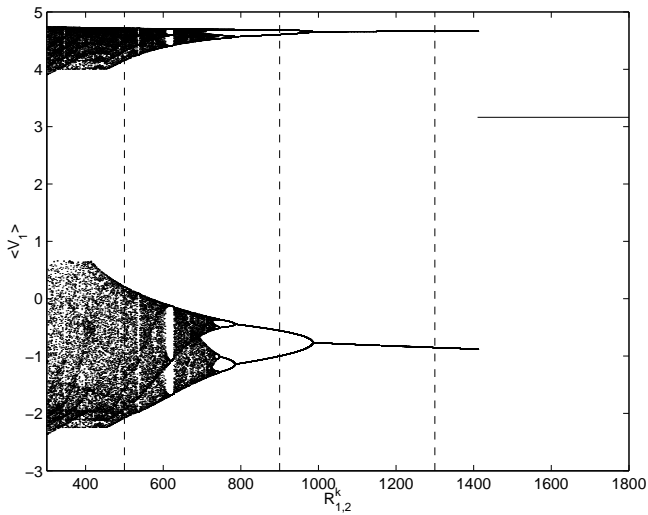,clip=,width=.48\textwidth}
  \caption{Bifurcation diagram of the reduced system
  \refeq{eq:popetau}. The Poincar\'e section of the attractor is shown
  as a function of the parameter mismatch between the two circuits,
  indicating that an increase in diversity induces a transition from a
  two-lobes to a one-lobe attractor. This macroscopic attractor then
  undergoes a backward period-doubling bifurcation cascade to a limit
  cycle, and eventually a fixed point is stabilized through a
  subcritical bifurcation. \label{fig:diagbifts}}
\end{figure}

The bifurcation diagram of the reduced system \eq{eq:popetau} can
be numerically computed for an interval of time scale differences
comparable to that accessible in our experimental setup. The
diagram provides the coherent
regimes of diffusively coupled Chua circuits as a function of
their parameter diversity and is indipendent of the population size.
As expected from the previous section,
for a low value of the coupling resistors (strong coupling), the
circuits are entrained on a chaotic attractor indistinguishable
from the uncoupled dynamics. However, when the coupling is reduced
a cascade of coherent regimes bifurcates from the chaotic
attractor and eventually the state of oscillator death is reached
(\reffig{fig:diagbifts}).

\begin{figure} \center
\epsfig{file=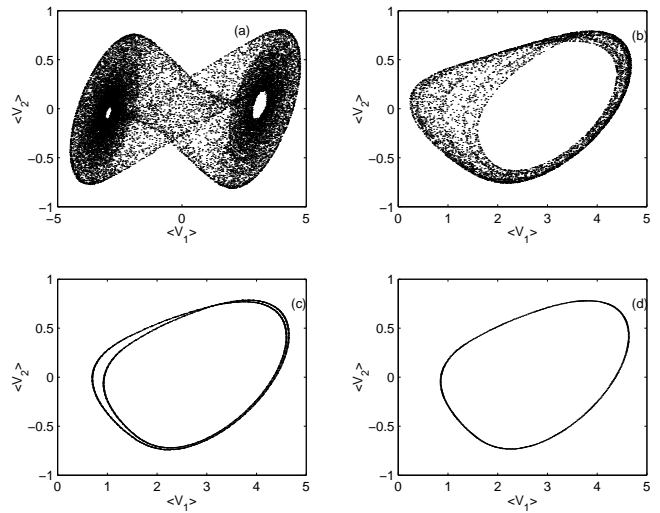,clip=,width=.48\textwidth}
\caption{Experimental behaviour of the coupled circuits at maximum
  coupling $R_k=0$ (a) and at values indicated in the bifurcation
  diagram of the reduced system Fig.\ \ref{fig:diagbifts}
  (b-d).\label{fig:diagbifexp}}
\end{figure}

The resistances $R^1_k=R^2_k=R_k$ of the coupled circuits can now
be tuned in order to target a specific collective regime,
according to the numerical characterization of the reduced system.
Figure\ \ref{fig:diagbifexp} shows some of the emergent dynamics
experimentally observed. The oscillation death regime also appears
in the circuits for $R^k$ slightly above $1.5$  k$\Omega$ (not
shown).

\section{Conclusions}
\label{sec:disc}

Although the importance of emergent behavior in populations of
nonidentical dynamical systems is widely recognized, only few
experimental systems offer the necessary degree of control for
testing the theoretical results. Here, we have proposed the
implementation of a circuit aimed at mimicking the coherent
behavior of populations with parameter mismatch in the region of
strong coupling. Theoretical results and numerical simulations
suggest that this case is representative of larger populations and
constitute a tool for addressing diversity-dependent coherent
dynamics.

The experimental system was composed of two symmetrically coupled
Chua circuits, the diversity in their components introduces a
mismatch in the internal parameters. We have shown that the
average dynamics of such coupled systems can be very different
from that of the individual circuits when decoupled. The order
parameter expansion developed for populations of strongly coupled
dynamical systems allowed us to predict such a collective behavior
based on the knowledge of the uncoupled dynamics. In this way, the
emergent dynamics can be tuned, by controlling the parameter
mismatch, onto an attractor that might not be reachable with any
of the individual circuits. The possibility of robustly targeting
a specific coherent regime can find a wide applicability in
populations of globally coupled non-identical systems, the chaotic
behavior of Chua circuits being just one example of individual
dynamics.

As a second example of diversity-induced qualitative changes of the
collective behavior, we have addressed the case when the circuits differ by
their time scales. In this case, the assumption that the coupling matrix is
independent of the parameters is valid for sufficiently small parameter
mismatch.

The coherency of the dynamics is not only the basic assumption that
allows us to analytically address the collective regimes, but also guarantees
that the results obtained for two circuits can be extended to populations of
many elements with a comparable variance of the parameter diversity. The
robustness of strong coupling regimes to changes in the number of interacting
circuits is confirmed by numerical simulations. These show indeed that the
collective behaviour is selected according to the parameter distribution
variance rather than from the actual population size.

\section{Acknowledgments} S.D.M. is thankful for the hospitality of DFI-IMEDEA
and is supported by the EIF Marie-Curie fellowship 010169. The authors thank W.
Korneta for interesting discussions and M. Oprandi for technical help. This
work has been partly supported by MEC (Spain) and FEDER through  project
CONOCE2 (FIS2004-00953).

\section{appendix}
The equations approximating the average dynamics of a population with time
scale mismatch can be computed for the Chua circuits that we have
addressed in this paper starting from \refeq{eq:chua}.
The collective dynamics is described by a system of six equations
(three for the variables of the average $X=[\nsum{v_1},\nsum{v_2},\nsum{cr}]$
and three for the shape parameter $W$. This can
be compactly written as \eqref{eq:tauj}, where $f(X)$ is a vector of
components:
\begin{eqnarray*}
f(1)=&&-\frac {G^b}
{C^1}\,v_1+\frac{v_2-v_1}{R\,C^1}-\frac{G^a-G^b}{2\,C^1}\\&&\times
\left(|v_1+B^p|-|v_1-B^p|\right)\\
f(2)=&&\frac {1} {C_2}\,\left(I+\frac{v_1-v_2}{R}\right)\\
f(3)=&&-\frac {1} {L}\,\left(I\,r^b-v_2\right).
\end{eqnarray*}

The coupling matrix $\mathcal K$ is:
\begin{eqnarray*}
\mathcal K=\left[
\begin{matrix}
& \frac{2}{C_1\,R^k_1} & 0 & 0\\
& 0 & \frac{2}{C_2\,R^k_2} & 0\\
& 0 & 0 & 0
\end{matrix}
\right]
\end{eqnarray*}
and the Jacobian matrix $\mathcal J$ is:
\begin{eqnarray*}
\mathcal J=\left[
\begin{matrix}
& j_{1,1} & \frac{1}{C_1\,R} &0\\\\
& \frac{1}{C_2\,R} & -\frac{1}{C_2\,R} & \frac 1 {C_2}\\\\
& 0 & -\frac 1 L & -\frac{r_b}{L}
\end{matrix}
\right],
\end{eqnarray*}
where
\begin{eqnarray*}
j_{1,1}=&& -\frac{G_b}{C_1}+\frac 1 {R\,C_1} +\frac {G_a-G_b}{2\,C_1}\\
&&\times \left[sgn(v_1+B_p)-sgn(v_1-B_p) \right].
\end{eqnarray*}

\bibliography{noise,synch}

\begin{thebibliography}{19}
\expandafter\ifx\csname natexlab\endcsname\relax\def\natexlab#1{#1}\fi
\expandafter\ifx\csname bibnamefont\endcsname\relax
  \def\bibnamefont#1{#1}\fi
\expandafter\ifx\csname bibfnamefont\endcsname\relax
  \def\bibfnamefont#1{#1}\fi
\expandafter\ifx\csname citenamefont\endcsname\relax
  \def\citenamefont#1{#1}\fi
\expandafter\ifx\csname url\endcsname\relax
  \def\url#1{\texttt{#1}}\fi
\expandafter\ifx\csname urlprefix\endcsname\relax\def\urlprefix{URL }\fi
\providecommand{\bibinfo}[2]{#2}
\providecommand{\eprint}[2][]{\url{#2}}

\bibitem[{\citenamefont{Pikovsky and Maistrenko}(2003)}]{pikovsky03}
\bibinfo{editor}{\bibfnamefont{A.}~\bibnamefont{Pikovsky}} \bibnamefont{and}
  \bibinfo{editor}{\bibfnamefont{Y.}~\bibnamefont{Maistrenko}}, eds.,
  \emph{\bibinfo{title}{Synchronization: Theory and application}}
  (\bibinfo{publisher}{Kuwler}, \bibinfo{address}{Dordrecht/Boston/London},
  \bibinfo{year}{2003}).

\bibitem[{\citenamefont{Mosekilde et~al.}(2002)\citenamefont{Mosekilde,
  Maistrenko, and Postnov}}]{mosekilde02}
\bibinfo{author}{\bibfnamefont{E.}~\bibnamefont{Mosekilde}},
  \bibinfo{author}{\bibfnamefont{Y.}~\bibnamefont{Maistrenko}},
  \bibnamefont{and} \bibinfo{author}{\bibfnamefont{D.}~\bibnamefont{Postnov}},
  \emph{\bibinfo{title}{{Chaotic Synchronization: Applications to Living
  Systems}}} (\bibinfo{publisher}{World Scientific},
  \bibinfo{address}{Singapore}, \bibinfo{year}{2002}).

\bibitem[{\citenamefont{Manrubia et~al.}(2004)\citenamefont{Manrubia,
  Mikhailov, and Zanette}}]{mikhailov04}
\bibinfo{author}{\bibfnamefont{S.~C.} \bibnamefont{Manrubia}},
  \bibinfo{author}{\bibfnamefont{A.~S.} \bibnamefont{Mikhailov}},
  \bibnamefont{and} \bibinfo{author}{\bibfnamefont{D.~H.}
  \bibnamefont{Zanette}}, \emph{\bibinfo{title}{Emergence of dynamical order:
  synchronization phenomena in complex systems}} (\bibinfo{publisher}{World
  Scientific}, \bibinfo{address}{Singapore}, \bibinfo{year}{2004}).

\bibitem[{\citenamefont{Oliva and Strogatz}(2001)}]{oliva01}
\bibinfo{author}{\bibfnamefont{R.~A.} \bibnamefont{Oliva}} \bibnamefont{and}
  \bibinfo{author}{\bibfnamefont{S.~H.} \bibnamefont{Strogatz}},
  \bibinfo{journal}{Int. J. of Bif. and Chaos} \textbf{\bibinfo{volume}{11}},
  \bibinfo{pages}{2359} (\bibinfo{year}{2001}).

\bibitem[{\citenamefont{Kuramoto}(1984)}]{kuramotobook}
\bibinfo{author}{\bibfnamefont{Y.}~\bibnamefont{Kuramoto}},
  \emph{\bibinfo{title}{Chemical Oscillations, Waves and Turbulence}}
  (\bibinfo{publisher}{Springer, Berlin}, \bibinfo{year}{1984}).

\bibitem[{\citenamefont{Kiss et~al.}(2002{\natexlab{a}})\citenamefont{Kiss,
  Zhai, and Hudson}}]{kiss02c}
\bibinfo{author}{\bibfnamefont{I.~Z.} \bibnamefont{Kiss}},
  \bibinfo{author}{\bibfnamefont{Y.}~\bibnamefont{Zhai}}, \bibnamefont{and}
  \bibinfo{author}{\bibfnamefont{J.~L.} \bibnamefont{Hudson}},
  \bibinfo{journal}{Science} \textbf{\bibinfo{volume}{296}},
  \bibinfo{pages}{1676} (\bibinfo{year}{2002}{\natexlab{a}}).

\bibitem[{\citenamefont{Dan{\o} et~al.}(1999)\citenamefont{Dan{\o},
  S{\o}rensen, and Hynne}}]{dano99}
\bibinfo{author}{\bibfnamefont{S.}~\bibnamefont{Dan{\o}}},
  \bibinfo{author}{\bibfnamefont{P.~G.} \bibnamefont{S{\o}rensen}},
  \bibnamefont{and} \bibinfo{author}{\bibfnamefont{F.}~\bibnamefont{Hynne}},
  \bibinfo{journal}{Nature} \textbf{\bibinfo{volume}{402}},
  \bibinfo{pages}{320} (\bibinfo{year}{1999}).

\bibitem[{\citenamefont{Winfree}(2001)}]{winfree01}
\bibinfo{author}{\bibfnamefont{A.~T.} \bibnamefont{Winfree}},
  \emph{\bibinfo{title}{The Geometry of Biological Time, 2nd edition}}
  (\bibinfo{publisher}{Springer}, \bibinfo{address}{New York},
  \bibinfo{year}{2001}).

\bibitem[{\citenamefont{Osipov and Sushik}(1998)}]{osipov98}
\bibinfo{author}{\bibfnamefont{G.}~\bibnamefont{Osipov}} \bibnamefont{and}
  \bibinfo{author}{\bibfnamefont{M.}~\bibnamefont{Sushik}},
  \bibinfo{journal}{Phys. Rev. E} \textbf{\bibinfo{volume}{58}},
  \bibinfo{pages}{7198} (\bibinfo{year}{1998}).

\bibitem[{\citenamefont{Matthews et~al.}(1991)\citenamefont{Matthews, Mirollo,
  and Strogatz}}]{matthews91}
\bibinfo{author}{\bibfnamefont{P.~C.} \bibnamefont{Matthews}},
  \bibinfo{author}{\bibfnamefont{R.~E.} \bibnamefont{Mirollo}},
  \bibnamefont{and} \bibinfo{author}{\bibfnamefont{S.~H.}
  \bibnamefont{Strogatz}}, \bibinfo{journal}{Physica D}
  \textbf{\bibinfo{volume}{52}}, \bibinfo{pages}{293} (\bibinfo{year}{1991}).

\bibitem[{\citenamefont{Rosenblum et~al.}(1996)\citenamefont{Rosenblum,
  Pikovsky, and Kurths}}]{rosenblum96}
\bibinfo{author}{\bibfnamefont{M.~G.} \bibnamefont{Rosenblum}},
  \bibinfo{author}{\bibfnamefont{A.}~\bibnamefont{Pikovsky}}, \bibnamefont{and}
  \bibinfo{author}{\bibfnamefont{J.}~\bibnamefont{Kurths}},
  \bibinfo{journal}{Phys. Rev. Lett.} \textbf{\bibinfo{volume}{76}},
  \bibinfo{pages}{1804} (\bibinfo{year}{1996}).

\bibitem[{\citenamefont{Kuramoto}(1975)}]{kuramoto75}
\bibinfo{author}{\bibfnamefont{Y.}~\bibnamefont{Kuramoto}}, in
  \emph{\bibinfo{booktitle}{International {S}ymposium on {M}athematical
  {P}roblems in {T}heoretical {P}hysics}}, edited by
  \bibinfo{editor}{\bibfnamefont{H.}~\bibnamefont{Araki}}
  (\bibinfo{publisher}{Springer, New York}, \bibinfo{year}{1975}),
  vol.~\bibinfo{volume}{39} of \emph{\bibinfo{series}{Lecture {N}otes in
  {P}hysics}}, p. \bibinfo{pages}{420}.

\bibitem[{\citenamefont{Kiss et~al.}(2002{\natexlab{b}})\citenamefont{Kiss,
  Zhai, and Hudson}}]{kiss02}
\bibinfo{author}{\bibfnamefont{I.~Z.} \bibnamefont{Kiss}},
  \bibinfo{author}{\bibfnamefont{Y.}~\bibnamefont{Zhai}}, \bibnamefont{and}
  \bibinfo{author}{\bibfnamefont{J.~L.} \bibnamefont{Hudson}},
  \bibinfo{journal}{Science} \textbf{\bibinfo{volume}{296}},
  \bibinfo{pages}{1676} (\bibinfo{year}{2002}{\natexlab{b}}).

\bibitem[{\citenamefont{{De Monte} and d'Ovidio}(2002)}]{demonte02}
\bibinfo{author}{\bibfnamefont{S.}~\bibnamefont{{De Monte}}} \bibnamefont{and}
  \bibinfo{author}{\bibfnamefont{F.}~\bibnamefont{d'Ovidio}},
  \bibinfo{journal}{Europhys. Lett.} \textbf{\bibinfo{volume}{58}},
  \bibinfo{pages}{21} (\bibinfo{year}{2002}).

\bibitem[{\citenamefont{{De Monte} et~al.}(2003)\citenamefont{{De Monte},
  d'Ovidio, and Mosekilde}}]{demonte03}
\bibinfo{author}{\bibfnamefont{S.}~\bibnamefont{{De Monte}}},
  \bibinfo{author}{\bibfnamefont{F.}~\bibnamefont{d'Ovidio}}, \bibnamefont{and}
  \bibinfo{author}{\bibfnamefont{E.}~\bibnamefont{Mosekilde}},
  \bibinfo{journal}{Phys. Rev. Lett.} \textbf{\bibinfo{volume}{90}},
  \bibinfo{pages}{054102} (\bibinfo{year}{2003}).

\bibitem[{\citenamefont{{De Monte} et~al.}(2004)\citenamefont{{De Monte},
  d'Ovidio, Chat\'e, and Mosekide}}]{demonte04a}
\bibinfo{author}{\bibfnamefont{S.}~\bibnamefont{{De Monte}}},
  \bibinfo{author}{\bibfnamefont{F.}~\bibnamefont{d'Ovidio}},
  \bibinfo{author}{\bibfnamefont{H.}~\bibnamefont{Chat\'e}}, \bibnamefont{and}
  \bibinfo{author}{\bibfnamefont{E.}~\bibnamefont{Mosekide}},
  \bibinfo{journal}{Phys. Rev. Lett.} \textbf{\bibinfo{volume}{92}},
  \bibinfo{pages}{254101} (\bibinfo{year}{2004}).

\bibitem[{\citenamefont{Ermentrout and Troy}(1989)}]{ermentrout89}
\bibinfo{author}{\bibfnamefont{G.~B.} \bibnamefont{Ermentrout}}
  \bibnamefont{and} \bibinfo{author}{\bibfnamefont{W.~C.} \bibnamefont{Troy}},
  \bibinfo{journal}{SIAM J. Math. Anal.} \textbf{\bibinfo{volume}{20}},
  \bibinfo{pages}{1436} (\bibinfo{year}{1989}).

\bibitem[{\citenamefont{Ermentrout}(1990)}]{ermentrout90}
\bibinfo{author}{\bibfnamefont{G.~B.} \bibnamefont{Ermentrout}},
  \bibinfo{journal}{Physica D} \textbf{\bibinfo{volume}{41}},
  \bibinfo{pages}{219} (\bibinfo{year}{1990}).

\bibitem[{\citenamefont{Atay}(2003)}]{atay03b}
\bibinfo{author}{\bibfnamefont{F.}~\bibnamefont{Atay}}, \bibinfo{journal}{Phys.
  Rev. Lett.} \textbf{\bibinfo{volume}{91}}, \bibinfo{pages}{094101}
  (\bibinfo{year}{2003}).

\end{thebibliography}

\end{document}